\documentclass[USenglish,twocolumn]{article}
\usepackage{color,soul}
\usepackage[utf8]{inputenc}				
\usepackage[big,online]{dgruyter}	
\usepackage{lmodern} 
\usepackage{microtype}
\usepackage[numbers,square,sort&compress]{natbib}
\usepackage{bm}
\usepackage{upgreek}
\theoremstyle{dgthm}

\theoremstyle{dgdef}

\begin{document}

	\articletype{Research Article}
	\received{Month	DD, YYYY}
	\revised{Month	DD, YYYY}
  \accepted{Month	DD, YYYY}
  \journalname{De~Gruyter~Journal}
  \journalyear{YYYY}
  \journalvolume{XX}
  \journalissue{X}
  \startpage{1}
  \aop
  \DOI{10.1515/sample-YYYY-XXXX}

\title{Generation of tunable optical skyrmions on Skyrme-Poincar\'e sphere}
\runningtitle{Tunable optical skyrmions on Skyrme-Poincar\'e sphere}

\author*[1]{Yijie Shen}
\author[2]{Eduardo Casas Mart\'inez} 
\author*[3,4]{Carmelo Rosales-Guzm\'an}
\runningauthor{Y.~Shen et al.}
\affil[1]{\protect\raggedright 
Optoelectronics Research Centre, University of Southampton, Southampton SO17 1BJ, United Kingdom, e-mail: y.shen@soton.ac.uk}
\affil[2]{\protect\raggedright 
Instituto Nacional de Astrof\'isica, \'Optica y Electr\'onica, Departamento de Óptica, Puebla, México}
\affil[3]{\protect\raggedright 
Centro de Investigaciones en Óptica, A.C., Loma del Bosque 115, Colonia Lomas del campestre, 37150 León, Gto., Mexico, e-mail: carmelorosalesg@cio.mx; and Wang Da-Heng Collaborative Innovation Center, Heilongjiang Provincial Key Laboratory of Quantum Manipulation and Control, Harbin University of Science and Technology, Harbin 150080, China, e-mail: carmelorosalesg@hrbust.edu.cn}
	
	
\abstract{{In recent time, the optical-analogous skyrmions, topological quasiparticles with sophisticated vectorial structures, have received an increasing amount of interest. Here we propose theortically and experimentally a generalized family of these, the tunable optical skyrmion, unveiling a new mechanism to transform between various skyrmionic topologies,} including N\'eel-, Bloch-, and anti-skyrmion types, via simple parametric tuning. {In addition, Poincar\'e-like geometric representation is proposed to visualize the topological evolution of tunable skyrmions, which we termed  Skyrme-Poincar\'e sphere, akin to the spin-orbit representation of complex vector modes}. To generate experimentally the tunable optical skyrmions we implemented a digital hologram system based on a spatial light modulator, showing great agreement with our theoretical prediction.}


\maketitle

\section{Introduction} 

Skyrmions are topologically protected quasiparticle in high-energy physics and condensed matter with salient vectorial textures~\cite{gobel2021beyond,fert2017magnetic}. This concept was recently {identified} by the optics and photonics community as a cuttin-edge topic. Recently, skyrmions were constructed in optical field as state-of-the-art optical structures and termed optical skyrmions~\cite{tsesses2018optical}. These were first generated in the electric field of evanescent wave~\cite{tsesses2018optical} followed by diverse forms constructed from different kinds of optical fields, such as, spin field of confined free-space waves~\cite{du2019deep,gutierrez2021optical}, Stokes vectors of paraxial vector beams~\cite{gao2020paraxial,lin2021}, magnetic vectors in propagating light pulses~\cite{shen2021supertoroidal}, and pseudospins in photonic crystals~\cite{karnieli2021emulating}. The creation of optical skyrmions has promising advanced applications {in fields} such as nanoscale metrology~\cite{dai2020plasmonic}, deep-subwavelength microscopy~\cite{du2019deep}, ultrafast vector imaging~\cite{davis2020ultrafast}, and topological Hall devices~\cite{karnieli2021emulating}, broadening the frontier of modern fundamental and applied physics.

{The advancement skyrmions provide is} mainly on their versatile topological textures, providing new degrees of freedom to shape vectorial fields and encode information. The skyrmionic configuration can be mapped into real-space magnetic materials and classified into diverse topologies~\cite{yu2010real,nagaosa2013topological}, including N\'eel-type~\cite{kezsmarki2015neel}, Bloch-type~\cite{gilbert2015realization}, and anti-skyrmion~\cite{nayak2017magnetic}. However, the control of diverse topologies of optical skyrmions {is an emerging topic  still its infancy}. As the first scheme of optical skyrmions, the evanescent electric fields on a plasmonic surface can only form N\'eel-type skyrmions~\cite{tsesses2018optical,davis2020ultrafast}. One year ago, a study of plasmonic skyrmion controlled between N\'eel- and Bloch-types was reported~\cite{bai2020dynamic}, but, soon after, such Bloch-type skyrmion was disproved~\cite{meiler2020dynamic,bai2020dynamicr}. The loophole-free observation of Bloch-type optical skyrmion was reported very recently in optical chiral multilayers~\cite{zhang2021bloch}. For the optical skyrmions in free space, Bloch-type skyrmion was proved in spin field of a tightly focused vortex beam~\cite{du2019deep}. And soon after, both N\'eel- and Bloch-type skyrmions were theoretically presented in Stokes vectors of paraxial vector beams~\cite{gao2020paraxial} and electric-spin fields in tightly focused vector beams~\cite{gutierrez2021optical}, however, no experimental results {have been reported for free-space skyrmions}. Finally, the optical anti-skyrmion also has never been proposed yet. 

In this paper, a closed-form expression {is} proposed to characterize a general class of optical skyrmions, where skyrmions with different textures (N\'eel-, Bloch-, and anti-skyrmion types) can be topologically transformed among each others via simple parametric tuning. Moreover, a novel {geometrical} model, Skyrme-Poincar\'e sphere, is proposed to universally map the topological transformation of tunable optical skyrmions, akin to the prior modal Poincar\'e sphere representing spin-orbital conversation. Importantly, we experimentally generate the tunable optical skyrmions with all the topological types in structured vector beams controlled by a digital hologram system, showing great agreement with our theoretical prediction.

\section{Theory} 
\subsection{Topologies of skyrmions} 
\begin{figure}[t!]
	\centering
	\includegraphics[width=\linewidth]{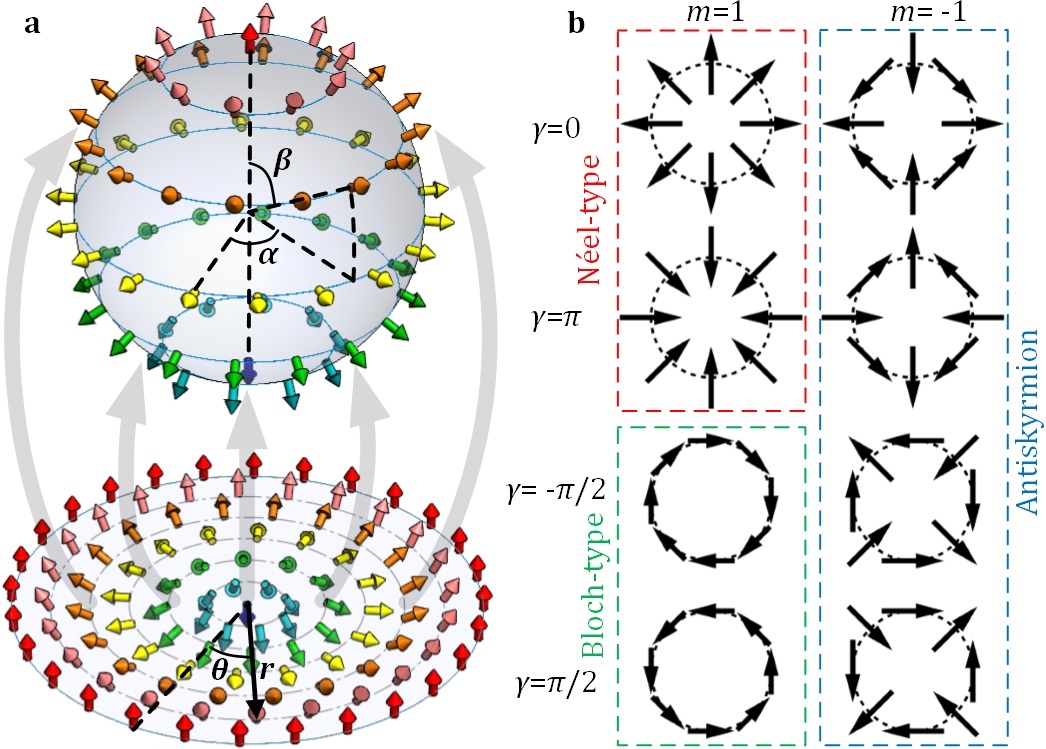}
	\caption{(a) The mapping from a normal skyrmion configuration, constructed by the colored arrows, to the unit sphere representation. (b) Some basic cases of the transversely projected vector ($E_x$ and $E_y$ components) distribution at a given radius of the skyrmions with various values of $m$ and $\gamma$.} 
	\label{f1}
\end{figure}

Topological properties of a skyrmionic configuration can be characterized by the skyrmion number defined by~\cite{nagaosa2013topological}:
\begin{equation}
    s=\frac{1}{4\pi }\iint_\sigma{\mathbf{n}\cdot \left( \frac{\partial \mathbf{n}}{\partial x}\times \frac{\partial \mathbf{n}}{\partial y} \right)}\text{d}x\text{d}y
\end{equation}
where $\mathbf{n}(x,y)$ represents the vector field to construct a skyrmion and $\sigma$ the region to confine the skyrmion, which can be infinity (for an isolated skyrmion) also can be a cell of a periodic distribution (for skyrmion lattices). The skyrmion number is an integer counting how many times the vector $\mathbf{n}(x,y)= \mathbf{n}(r\cos\theta, r\sin\theta)$ wraps around the unit sphere, as the mapping shown in Fig.~\ref{f1}(a). For mapping to the unit sphere, the vector can be given by $\mathbf{n}=(\cos\alpha (\theta )\sin\beta (r),\sin\alpha (\theta )\sin\beta (r),\cos\beta (r))$. Also, the skyrmion number can be separated into two integers: 
\begin{align}
\nonumber
s=&\frac{1}{4\pi }\int_{0}^{r_\sigma}{\text{d}r}\int_{0}^{2\pi }{\text{d}\theta }\frac{\text{d}\beta (r)}{\text{d}r}\frac{\text{d}\alpha (\theta )}{\text{d}\theta }\sin \beta (r)\\
=&\frac{1}{4\pi }[\cos\beta (r)]_{r=0}^{r=r_\sigma}[\alpha (\theta )]_{\theta =0}^{\theta =2\pi }=p\cdot m
\end{align}
the polarity, $p=\frac{1}{2}[\cos\beta (r)]_{r=0}^{r=r_\sigma}$, {indicates the direction of the vector} down (up) at center $r=0$ and up (down) at {boundary $r\to r_\sigma$} for $p=1$ ($p=-1$), and the vorticity, $m=\frac{1}{2\pi }[\alpha (\theta )]_{\theta =0}^{\theta =2\pi }$, controls the distribution of the transverse field components~\cite{wang2018theory}. In the case of a helical distribution, an initial phase $\gamma$ should be added, $\alpha (\theta )=m\theta +\gamma$. If we consider the transverse vector components at a given radius (or a given latitude angle $\beta$ in the unit-sphere representation), the $\gamma$ reveals the inclined angle of initial vector in the circular array, see Fig.~\ref{f1}(b). For the $m=1$ skyrmion, the cases of $\gamma=0$ and $\gamma=\pi$ are classified as N\'eel-type, and the cases of $\gamma=\pm\pi/2$ are Bloch-type. The case of $m=-1$ is anti-skyrmion. 
\begin{figure*}[t!]
	\centering
	\includegraphics[width=\linewidth]{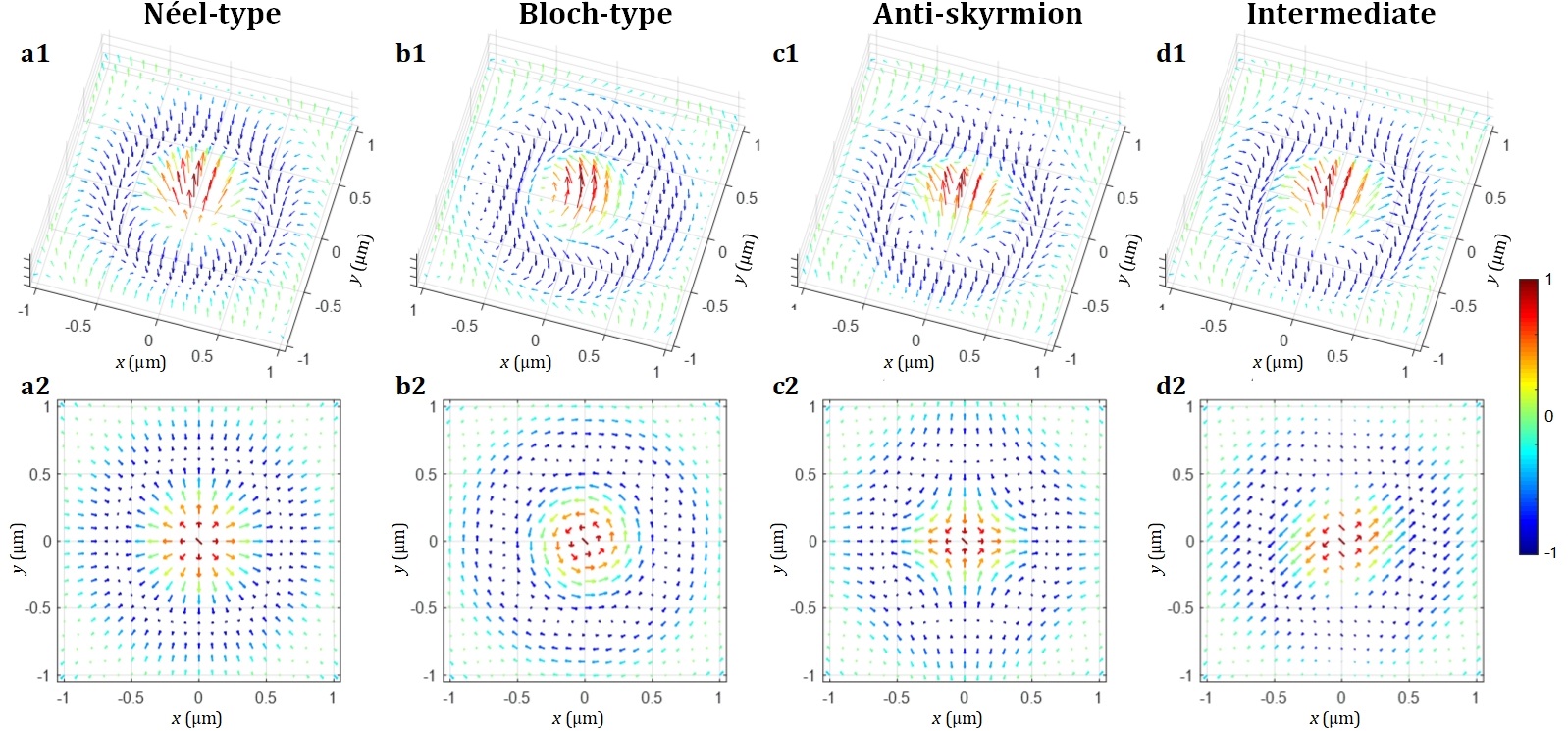}
	\caption{(a-d) The simulated results (a1-d1) The 3D vector distributions, {$(n_x,n_y,n_z)$} and (a2-d2) the transverse component vectors, {$(n_x,n_y)$}, for tunable optical skyrmions with diverse topological textures of N\'eel-type (a, $\phi_1=\phi_2=0$), Bloch-type (b, $\phi_1=\phi_2=\pi/2$), anti-skyrmion (b, $\phi_1=0$, $\phi_2=\pi$), and an intermediate state (b, $\phi_1=0$, $\phi_2=\pi/2$), respectively, and other Parameters used in simulation: $k_z=2\pi$~nm$^{-1}$, $k_\parallel=2k_z$ $N=3$, $\phi_1$ and $\phi_2$.} 
	\label{f2}
\end{figure*}

\subsection{Tunable optical skyrmions} {In order to derive a closed-form expression that allows to tune between optical skyrmions with diverse topological textures via simple parameters, we start by a classic model of optical skyrmion, which was also also the first model of optical skyrmion.} Namely, the skyrmionic vector field, $\mathbf{n}=(n_x,n_y,n_z)$, is constructed by the electric field vectors, $(E_x,E_y,E_z)^\text{T}/|\mathbf{E}|$, in a confined surface plasmon polaritons (SPP) wave~\cite{tsesses2018optical}:
\begin{align}
\mathbf{n}={{E}_{0}}{{\text{e}}^{-|{{k}_{z}}|z}}\sum\limits_{n=1}^{N}{\left( \begin{matrix}
   -\frac{|{{k}_{z}}|z\cos {{\theta }_{n}}}{{{k}_{\parallel }}}\sin \left( {{k}_{\parallel }} \mathbf{r}_{\parallel }\cdot {\bm{\uptheta }_{n}} \right)  \\
   -\frac{\left| {{k}_{z}} \right|\sin {{\theta }_{n}}}{{{k}_{\parallel }}}\sin \left( {{k}_{\parallel }} \mathbf{r}_{\parallel }\cdot {\bm{\uptheta }_{n}} \right)   \\
   \cos \left( {{k}_{\parallel }} \mathbf{r}_{\parallel }\cdot {\bm{\uptheta }_{n}} \right)   \\
\end{matrix} \right)}\label{n1}
\end{align}
where $E_0$ is a normalized amplitude, $k_\parallel$ and $k_z$ are transverse (in-plane) and axial wavenumbers, $\mathbf{r}_{\parallel }=(x,y)$ and $\bm{\uptheta }_{n}=(\cos {{\theta }_{n}},\sin {{\theta }_{n}})$. The physical meaning of Eq.~(\ref{n1}) is the superposition of $N$ standing-wave SPPs along directions with equally distributed in-plane angles $\theta_n$ ($n=1,2,\cdots,N$). For common circular-shape skyrmions, $N$ should large enough, ideally $N\to\infty$. For the case of skyrmion lattice, $N$ should be a integer related to the boundary geometry, e.g. $N=3$ is set for simulating hexagonal skyrmion-lattice field with $\theta_n=[-\pi/3,0,\pi/3]$~\cite{tsesses2018optical}. Figure~\ref{f2}(a) shows a simulated result of skyrmionic vector distribution using Eq.~(\ref{n1}). However, Eq.~(\ref{n1}) can only represent N\'eel-type skyrmion (lattice). Here we propose a mathematically generalized form {that breaks} this limit:
\begin{equation}
\mathbf{n}={{E}_{0}}{{\text{e}}^{-|{{k}_{z}}|z}}\sum\limits_{n=1}^{N}{\left( \begin{matrix}
   -\frac{\left| {{k}_{z}} \right|\cos ({{\theta }_{n}}+\phi_1)}{{{k}_{\parallel }}}\sin \left( {{k}_{\parallel }} \mathbf{r}_{\parallel }\cdot {\bm{\uptheta }_{n}} \right)   \\
   -\frac{\left| {{k}_{z}} \right|\sin ({{\theta }_{n}}+\phi_2)}{{{k}_{\parallel }}}\sin \left( {{k}_{\parallel }} \mathbf{r}_{\parallel }\cdot {\bm{\uptheta }_{n}} \right)   \\
   \cos \left( {{k}_{\parallel }} \mathbf{r}_{\parallel }\cdot {\bm{\uptheta }_{n}} \right)   \\
\end{matrix} \right)}\label{n2}
\end{equation}
Compared with Eq.~(\ref{n1}), Eq.~(\ref{n2}) has {two additional} parameters, $\phi_1$ and $\phi_2$, the control of {which} can drive the skyrmionic vector field to cover all the three types of topologies (N\'eel-, Bloch- and anti-skyrmion). For $\phi_1=\phi_2=\phi\in[0,2\pi]$, Eq.~(\ref{n2}) represents tunable skyrmion between N\'eel- ($\phi=0,\pi$) and Bloch-types ($\phi=\pi/2,3\pi/2$), see \textbf{Video~1} for the evolution movie and a result of the Bloch-type skyrmion, which is shown in Figs.~\ref{f2}(b). For $\phi_1=0$ and $\phi_2=\phi\in[0,2\pi]$, Eq.~(\ref{n2}) represents tunable skyrmion between N\'eel-type ($\phi=0$) and anti-skyrmion ($\phi=\pi$), see \textbf{Video~2} for the movie, a result of an anti-skyrmion in Figs.~\ref{f2}(c) and an interposed state in such transformation is shown in Figs.~\ref{f2}(d). When $\phi_1=\pi/2$ and $\phi_2=\phi\in[0,2\pi]$, Eq.~(\ref{n2}) represents tunable skyrmion between Bloch-type ($\phi=\pi/2$) and anti-skyrmion ($\phi=3\pi/2$), see \textbf{Video~3} for the movie.

\begin{figure}[t!]
	\centering
	\includegraphics[width=\linewidth]{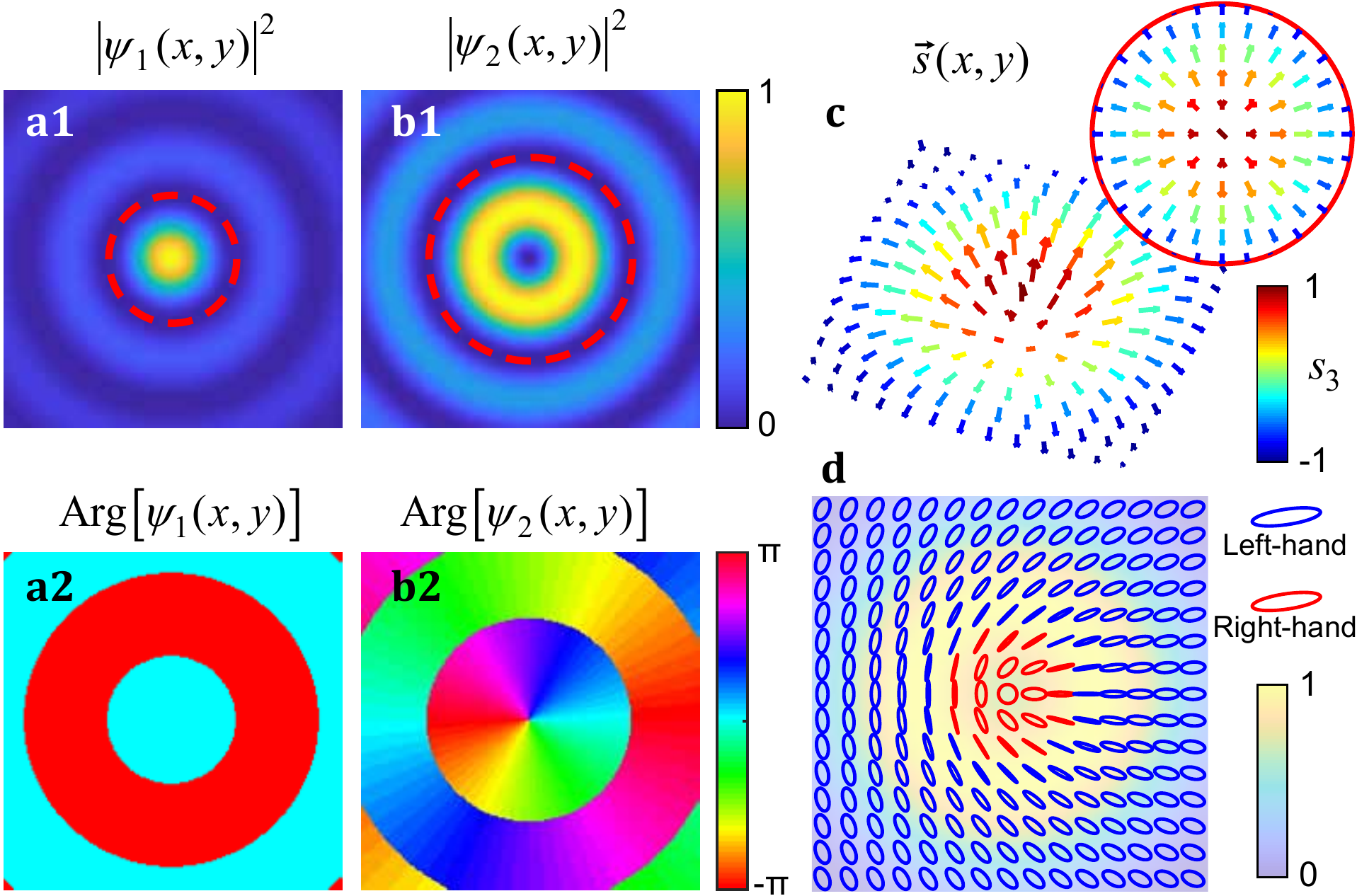}
	\caption{(a,b) Intensity (a1,b1) and phase (a2,b2) distributions of $\psi_1(x,y)$ (a) and $\psi_1(x,y)$ (b) for the case of $\phi_1=\phi_2=0$, the red-dash lines mark the effective regions to induce optical skyrmion. (c,d) The skyrmion in Stokes vector field with transverse component inset (c) and the intensity and polarization distributions (d) of the corresponding vector beam.} 
	\label{f3}
\end{figure}

The mathematical representation of Eq.~(\ref{n2}) is correct for tunable optical skyrmion, but its experimental realization is still cumbersome. Attempting to use SPP to generate tunable skyrmions, as described by Eq.~(\ref{n2}), the two parameters $\phi_1$ and $\phi_2$ would induce extreme conditions that are hard to fulfill, as a stable SPP field requires rigorous conditions than a free-space light field.
Importantly, recent advancements had made possible to successfully realize optical skirmions with different optical vector fields in both matter and free space~\cite{du2019deep,gutierrez2021optical,gao2020paraxial,karnieli2021emulating}. {These advancements allows to propose a practical scheme for the generation of tunable optical skyrmions in free space, via the Stokes vectors of structured vector beams.} The Stokes vector $\mathbf{s}=(s_1,s_2,s_3)$ can represent an arbitrary state of polarization, {as points on the surface of unit-radius sphere known as the Poincar\'e sphere~\cite{rosales2018review}. In spherical coordinates, the optical field $\bm{\uppsi }=\cos ({\theta }/{2}){e}^{-i\varphi/2}\mathbf{R}+\sin ({\theta }/{2}){{e}^{i\varphi/2}}\mathbf{L}$, where $\mathbf{R}$ and $\mathbf{L}$ represent right- and left-handed circularly polarized (RCP and LCP) eigenstates, respectively, is represented as $\mathbf{s}=(\cos \varphi \sin \theta ,\sin\varphi \sin \theta ,\cos \theta )$.} In order to construct skyrmions by Stokes vector, we need to first construct a vector beam with customized spatial modes of the form,
\begin{equation}
    \bm{\uppsi }(x,y)={{\psi }_{1}}(x,y)\mathbf{R}+{{\psi }_{2}}(x,y)\mathbf{L}.
    \label{vb}
\end{equation}
{The spatial modes $\psi_1$ and $\psi_2$ should have the same form of the tunable skyrmion, given by Eq.~(\ref{n2}), that is,}
\begin{align}
    {\psi }_{1}(x,y)&=\cos [{\theta (x,y)}/{2}]{e}^{-i{\varphi (x,y)}/{2}}\\
    {{\psi }_{2}}(x,y)&=\sin [{\theta (x,y)}/{2}]{{e}^{i{\varphi (x,y)}/{2}}},
\end{align}
with
\begin{align}
    \theta (x,y)&={{\cos }^{-1}}[{{n}_{z}}(x,y)]\\
    \varphi (x,y)&=\sin^{-1}\left[\frac{{{n}_{y}}(x,y)}{\sin \theta(x,y) }\right]={{\cos }^{-1}}\left[\frac{{{n}_{x}}(x,y)}{\sin \theta(x,y) }\right]
\end{align}
where, $(n_x,n_y,n_z)$ is exactly based on theoretical tunable skyrmion of Eq.~(\ref{n2}). We can now tailor vector beams to realize tunable skyrmions {via their} Stokes vector fields. {As a way of example, a numerical simulation of the N\'eel-type skyrmionic vector beam, for which $\phi_1=\phi_2=0$, is shown in Fig. ~\ref{f3}.} The spatial modes $\psi_1(x,y)$ and $\psi_2(x,y)$ are shown in Figs.~\ref{f3}(a,b), which share the same form of zero- and first-order Bessel modes, respectively. The simulation results also show that the effective regions to induce an skyrmionic structure are the central lobes of the zero- and first-order Bessel modes, as marked by red-dashed lines in Figs.~\ref{f3}(a,b). Importantly, the contribution of the outside rings, far away the skyrmion center, is almost negligible, which means, we can replace the zero- and first-order Bessel modes by the fundamental Gaussian and first-order Laguerre-Gaussian (LG) modes. The Stokes vector field of the corresponding vector beam given by Eq.~(\ref{vb}) shows a perfect N\'eel-type skyrmion, as evinced in Figs.~\ref{f3}(c,d).

\subsection{Skyrme-Poincar\'e sphere} 
\begin{figure}[t!]
	\centering
	\includegraphics[width=\linewidth]{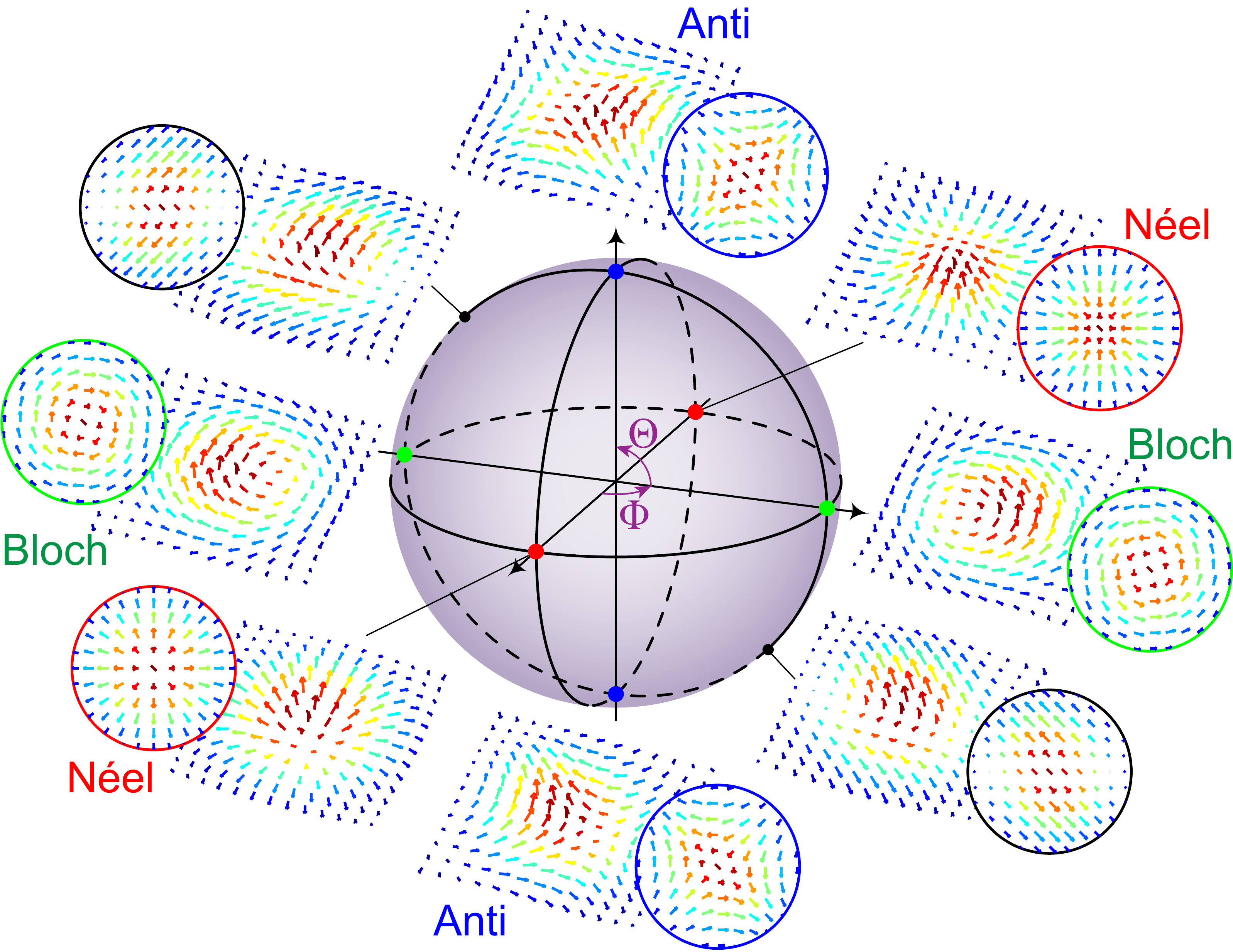}
	\caption{The Skyrme-Poincar\'e sphere to represent universal topological evolution of tunable optical skyrmion, with special topology as a point on it: Red points, N\'eel-type; green points, Bloch-type; Blue points (the poles), anti-skyrmion.} 
	\label{fps}
\end{figure}

In this section we present a graphical model, which we termed \textit{Skyrme-Poincar\'e sphere}, to depict the general topological transformation of tunable skyrmion onto a Poincar\'e-like sphere. Without loss of generality, setting $\phi_1=\Phi\in[0,2\pi]$ and $(\phi_2-\phi_1)/2=\Theta\in[-\pi/2,\pi/2]$, we can regard $\Phi$ and $\Psi$ as angles of longitude and latitude, respectively, to map the corresponding topological skyrmion to a corresponding point on the Skyrme-Poincar\'e sphere, as shown in Fig.~\ref{fps}. {For $\Theta=0$, the skyrmion transforms between N\'eel ($\Phi=0,\pi$) and Bloch ($\Phi=\pm\pi/2$), which are represented on the equator. The anti-skyrmions are located at the poles, for $\Theta=0$ and $\Theta=\pm\pi/2$, which are independent of $\Phi$.} This topology is extremely similar to the classic Poincar\'e sphere or modal Poincar\'e sphere that maps the evolution of spin/polarization or orbital angular momentum (OAM) of light~\cite{padgett1999poincare,dennis2017swings,shen20202}, where the equator represents the non-spin or non-OAM modes, while the route from equator to the poles corresponds to the introducing of spin or OAM charge. Similarly in Skyrme-Poincar\'e sphere, the equatorial transformation is skyrmion-number-conserved, e.g., for the normal polarity $p=1$, the skyrmion number is always $s=1$ for the tunable skyrmion between N\'eel and Bloch. While, the area from equator to the poles reveals the transformation to anti-skyrmion ($s=-1$) introducing a fractional skyrmion-number evolution, akin to the fractional OAM evolution from equation to poles on a Poincar\'e sphere.

The Skyrme-Poincar\'e sphere has more sophisticated relation to OAM Poincar\'e sphere under the skyrmionic vector beam representation. Based on the methods introduced in last section, we can derive the parametric form of corresponding vector beam dependent on $\Phi$ and $\Theta$:
\begin{equation}
    \bm{\Psi }=\Psi_0\mathbf{R}+\underbrace{\left[{\cos ({\Theta }/{2}){e}^{-i\Phi/2}\Psi_{-1}+\sin ({\Theta }/{2}){{e}^{i\Phi/2}}\Psi_{1}}\right]}_{\text{OAM Poincar\'e sphere}}\mathbf{L}
\end{equation}
where $\Psi_0$ is zero-order Bessel (or fundamental Gaussian) mode and $Psi_{\pm1}$ are $\pm1$th-order Bessel modes (or LG modes) carrying opposite OAM with $\pm1$ topological charges. And the spatial mode of LCP component is exactly the form of conventional OAM Poincar\'e sphere driven by the longitude and latitude angles $\Phi$ and $\Theta$, revealing the OAM conversion between $\pm1$ topological charges. Therefore the Skyrme-Poincar\'e sphere representing skyrmionic topological transformation can be bidirectionally mapped to conventional OAM Poincar\'e sphere representing OAM transformation. As such, many Poincar\'e-related applications, such as spin-orbit conversion, geometric phase transition, encoding and communication~\cite{shen2019optical,dennis2017swings,rosales2018review}, are expected to transfer to optical skyrmion.

\section{Experiment} 

\begin{figure}[tb]
	\centering
	\includegraphics[width=\linewidth]{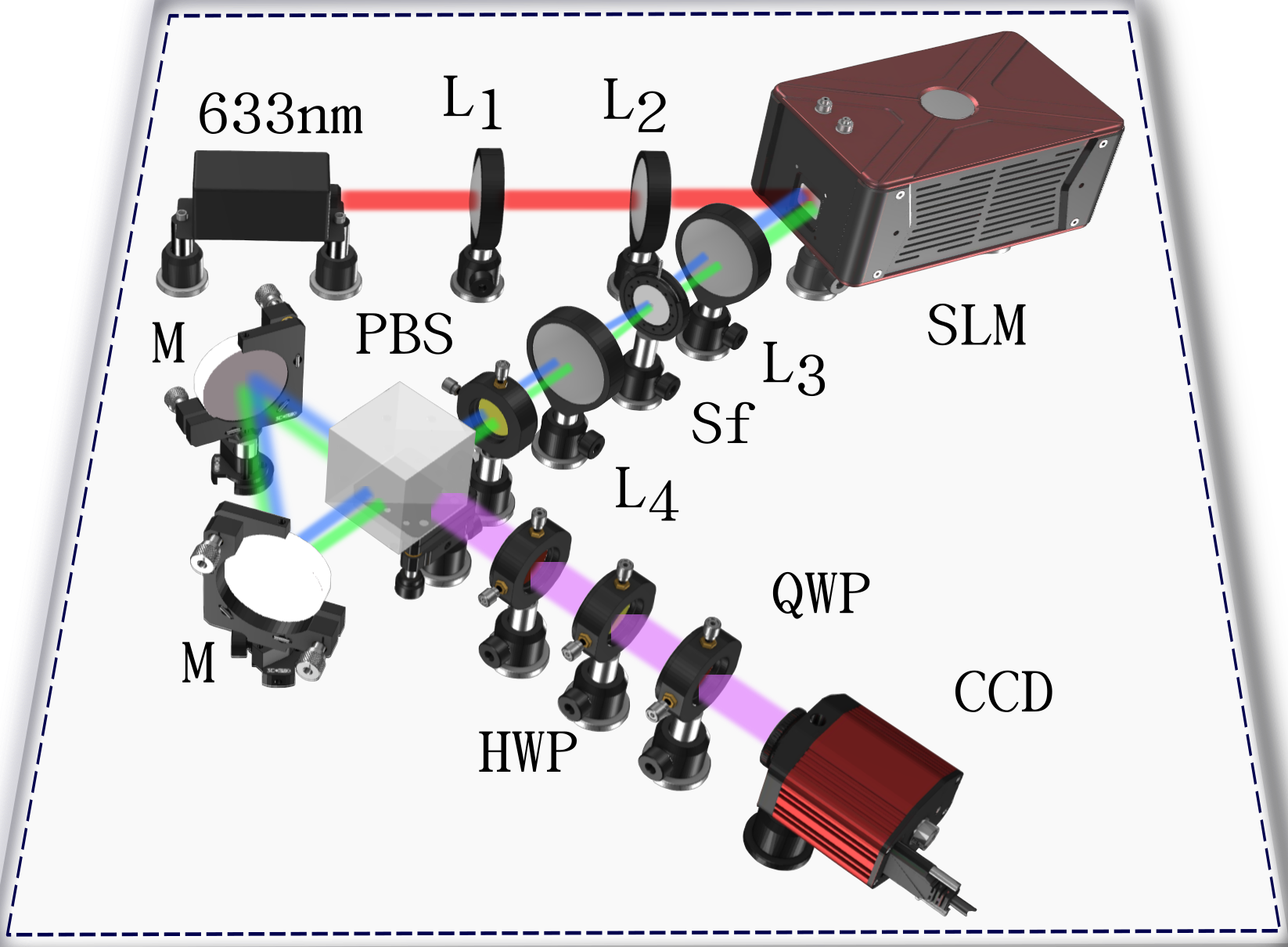}
	\caption{Schematic representation of the optical setup implemented to generate tunable optical skyrmions. L$_1$ -- L$_4$:lenses, PBS: Polarizing Beam Splitter, SLM:Spatial Light Modulator HWP: Half-Wave plate, QWP: Quarter-Wave Plate, M: Mirror, SF: Spatial Filter, CCD: Charge Coupled Device Camera} 
	\label{setup}
\end{figure}

In order to realize experimentally the tunable optical skyrmions, we implemented a highly stable optical setup capable to generate arbitrary optical beams with almost any state of polarisation and spatial profile \cite{PerezGarcia2017}, which is schematically shown in Fig. \ref{setup} . Such optical setup comprises the use of a Spatial Light Modulator (SLM) for arbitrary control of the spatial distribution of the optical field \cite{SPIEbook}. The setup starts with a horizontally polarised HeNe (632.8 nm) laser beam, collimated and expanded to fully cover the liquid crystal screen of the SLM. The screen of the SLM is digitally split into two independent screens, each of which is addressed with an independent digital hologram that generates, in the first diffraction order, two independent optical fields. Each digital hologram is superimposed with a linear grating to separate the different diffraction orders. Afterwards, the first diffraction order is spatially filtered using a telescope formed by two lenses and a spatial filter. Afterwards, the two optical fields are redirected to a common-path triangular Sagnac interferometer comprising a Polarising Beam Splitter (PBS) and two mirrors. Prior to entering the interferometer, both beams are rotated to a diagonal polarisation state. In this way, when both beams enter the interferometer, after traversing the PBS, each of them is separated into two new beams with orthogonal linear polarisation states, horizontal and vertical, traveling along opposite optical paths. After a round trip, all four beams exit the interferometer from the opposite side of the PBS, two with horizontal polarization and two with vertical. Finally, the horizontal polarisation component of one of the beams y aligned co-axially with the vertical polarisation component of the other beam. To ensure a perfect coaxial superposition, a fine tuning is performed digitally by adjusting the period of the linear gratings encoded on each digital hologram. 

\begin{figure}[tb]
	\centering
	\includegraphics[width=.98\linewidth]{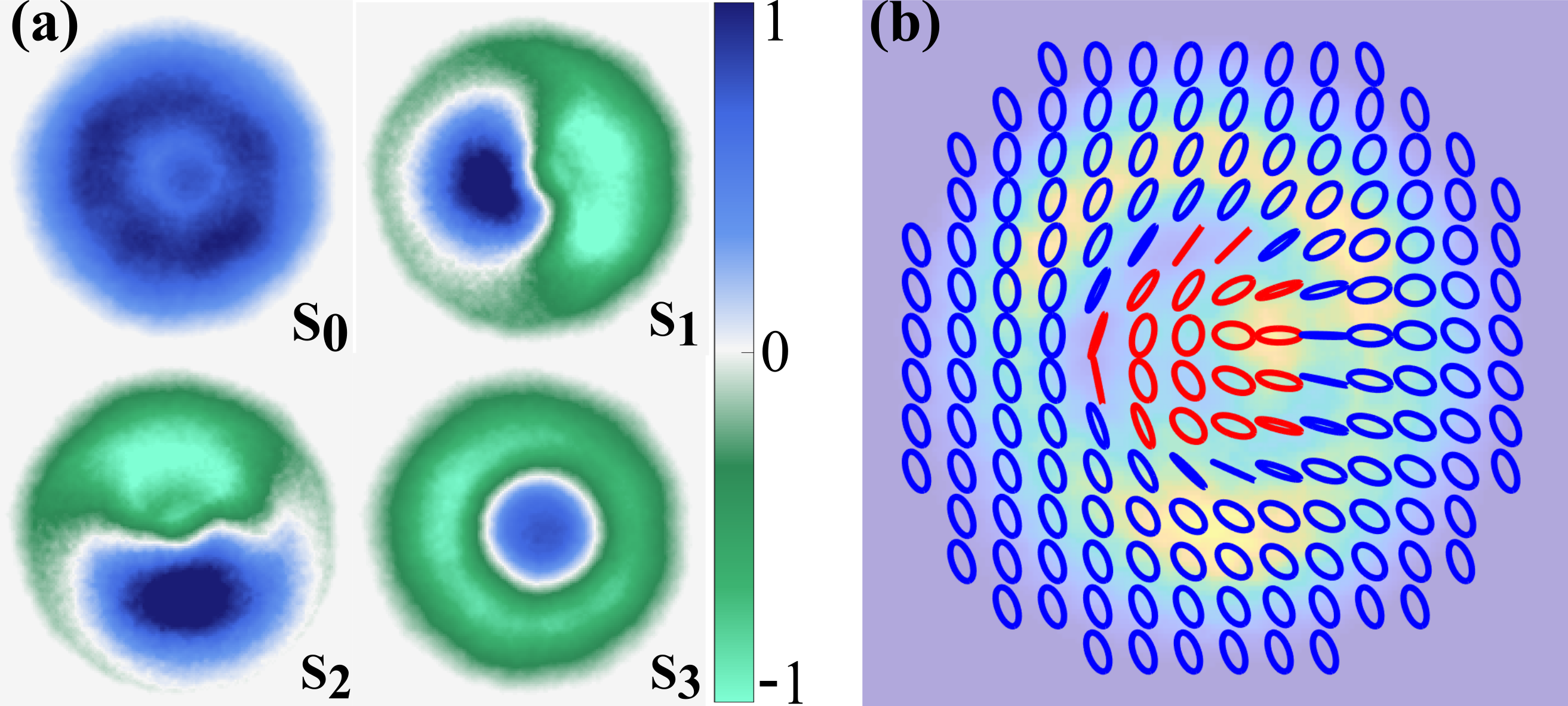}
	\caption{Experimentally reconstructed (a) Stokes parameters and (b) intensity distribution of the vector field overlapped with its corresponding polarisation distribution.}
	\label{stokes}
\end{figure}
\begin{figure*}[tb]
	\centering
	\includegraphics[width=\linewidth]{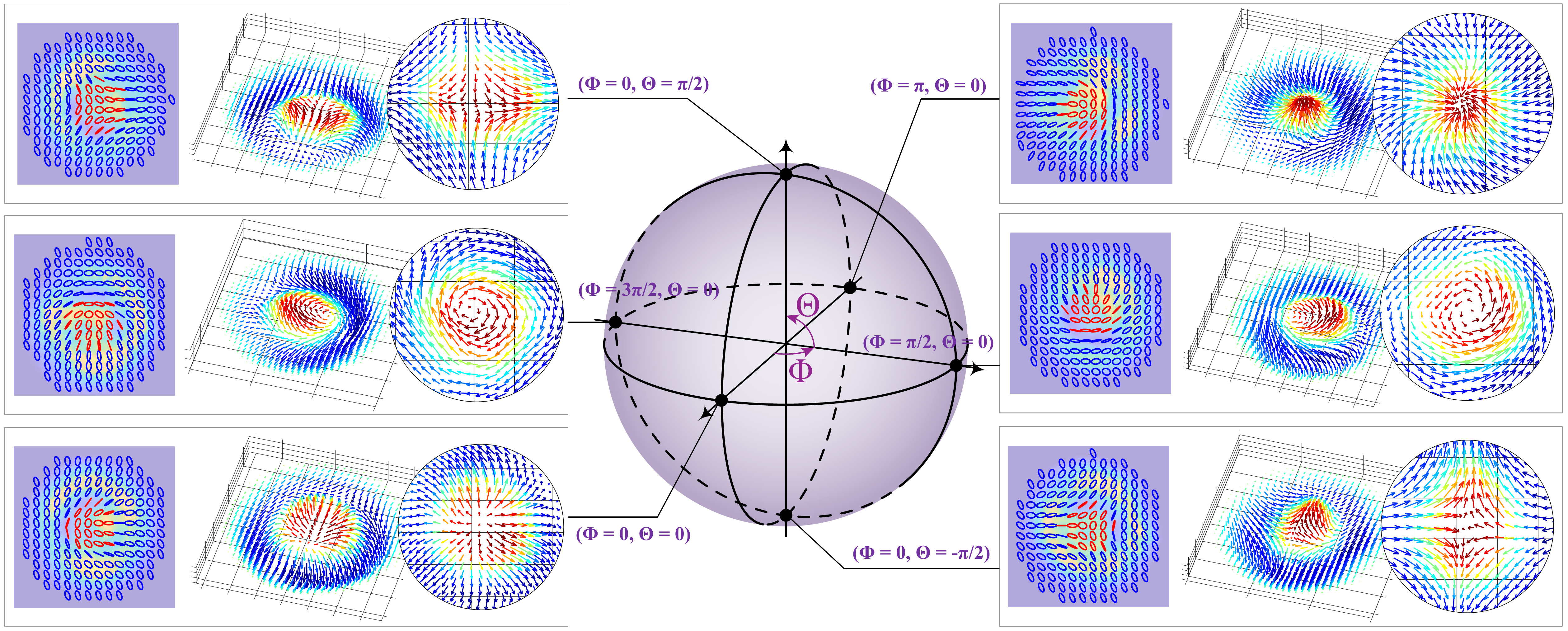}
	\caption{Experimentally results of tunable Stokes skyrmions with controlled topological textures corresponding to selective points on Skyrme-Poincar\'e sphere (see gray boxes), each box include the measured intensity and polarization distribution of the skyrmionic beam, the vector distribution of the Stokes skyrmions, and the zoom-in planform of the vector field for clearly distinguishing the topological texture.}
	\label{exp}
\end{figure*}

The experimental reconstruction of the optical skyrmions was achieve through Stokes polarimetry, more specifically, through the reconstruction of the Stokes parameters $S_0$, $S_1$, $S_2$ and $S_3$, which were computed from a set of intensity measurements. To this end, a second stage was build at the output port of the interferometer, where a series of phase retarders and a Charge-Coupled Device (CCD) camera allowed to measure the required intensities, from which the Stokes parameters where reconstructed using the relations,
\begin{equation}\label{Eq:Stokes}
\begin{split}
\centering
     &S_{0}=I_{H}+I_{V},\hspace{19mm}
     S_{1}=2I_{H}-S_{0},\\
     &S_{2}=2I_{D}-S_{0},\hspace{19mm}
     S_{3}=2I_{R}-S_{0},
\end{split}
\end{equation} 
where, $I_H$, $I_V$, $I_D$ and $I_R$ represent the intensity of the horizontal, vertical, diagonal and right-handed polarisation components, respectively. To experimentally measure $I_H$, $I_V$ and $I_D$ we passed the optical field through a linear polariser set at $0^\circ$, $45^\circ$ and $95^\circ$, respectively, while the intensity of the $I_R$ polarisation component was acquired by passing it simultaneously through a QWP at $45^\circ$ and a linear polariser at $90^\circ$ (see \cite{Rosales2020} for further details). Figure \ref{stokes}(a) show an example of the experimental Stokes parameters along with the intensity profile overlapped with its corresponding polarization distribution, Fig.~\ref{stokes}(a), for the specific case $\phi_1=\phi_2=0$.

With our digital hologram system, we realized a controlled generation of Stokes skyrmions with arbitrary topological textures on the Skyrme-Poincar\'e sphere. Figure~\ref{exp} shows our experimentally generated LG-based results of tunable Stokes skyrmions on typical points of Skyrme-Poincar\'e sphere, including the measured intensity and polarization distributions, Stokes vector fields, and the zoom-in planforms of the vector field for clearly distinguishing the topological textures. See more experiential results in Supplementary materials (Video~4 and Video~5 for the LG mode and Bessel mode based results).

\section{Discussion} 
Although the generalized tunable optical skyrmion model is inspired from the case of SPP electric field, and realized by Stokes vector field, it is not limited in these cases and can be freely extendable. Because the new concept of tunable skyrmion has a universal mathematical parametrization for generalized topology, the vector can refer to other kinds of optical field of structured light for the further exploration. For example, we can also use spin vectors in a tightly focused beam~\cite{du2019deep,gutierrez2021optical}, electric or magnetic vectors in a propagating structured pulse~\cite{shen2021supertoroidal}, and pseudospins in nonlinear media~\cite{karnieli2021emulating}. It is also a exciting direction to create more quasiparticle with complex topological states beyond skyrmions, such as skyrmion bags~\cite{foster2019two} and skyrmion tubes~\cite{kuratsuji2021evolution}, meron and bimeron~\cite{shen2021topological,krol2021observation,guo2020meron}, into structured light. Even though in this work, we used an SLM to generate skyrmionic beams, Digital Mirror Devices represent an alternative means to generated them, which present several advantages over SLMs, such as, their high refresh rates, polarisation independence and low cost\cite{Rosales2020}. 

Conclusively, we proposed an extended family of tunable optical skyrmion, enabling flexible transformation among various kinds of skyrmionic topological textures. A graphical model, Skyrme-Poincar\'e sphere, is proposed to universally represent the topological evolution of tunable optical skyrmion. We experimentally generated such tunable optical skyrmions in Stokes vector fields of customized vector beams from a well-designed digital hologram system. This is the first-know experimental generation of topology-tunable optical skyrmions in free space, also the first realization of optical anti-skyrmions. Our methods provide a new platform for optical information storage, communication, and cryptography using skyrmionic topological states of light.


\begin{funding}
  National Natural Science Foundation of China (61975047).
\end{funding}

\bibliography{sample}
\bibliographystyle{unsrtnat}
\end{document}